\documentclass[aps,prl,twocolumn,groupedaddress,showpacs]{revtex4-1}
\usepackage{graphicx}
\usepackage{latexsym}
\usepackage{amsmath}
\usepackage{txfonts}
\usepackage{helvet}
\usepackage{braket}
\usepackage{amscd}
\usepackage{hyperref}

\begin{document}

\title{Pressure-Tuned Exchange Coupling of a Quantum Spin Liquid in the Molecular Triangular Lattice $\kappa$-(ET)$_2$Ag$_2$(CN)$_3$} 

\author{Yasuhiro Shimizu,$^{1}$ Takaaki Hiramatsu,$^{2}$ Mitsuhiko Maesato,$^{3}$ Akihiro Otsuka,$^{3,4}$ Hideki Yamochi,$^{3,4}$ Akihiro Ono,$^{1}$ Masayuki Itoh,$^{1}$ Makoto Yoshida,$^{5}$ Masashi Takigawa,$^{5}$ Yukihiro Yoshida,$^{2}$ Gunzi Saito$^{2,6}$\\}
\affiliation{$^{1}$Department of Physics, Nagoya University, Chikusa, Nagoya 464-8602, Japan.}
\affiliation{$^{2}$Faculty of Agriculture, Meijo University, Tempaku-ku, Nagoya 468-8502, Japan.}
\affiliation{$^{3}$Division of Chemistry, Graduate School of Science, Kyoto University, Sakyo-ku, Kyoto 606-8502, Japan.}
\affiliation{$^{4}$Research Center for Low Temperature and Materials Sciences, Kyoto University, Sakyo-ku, Kyoto 606-8501, Japan.}
\affiliation{$^{5}$Institute for Solid State Physics, University of Tokyo, Kashiwa, Chiba 277-8581, Japan.}
\affiliation{$^{6}$Toyota Physical and Chemical Research Institute, Nagakute, Aichi 480-1192, Japan.}

\date{\today}

\begin{abstract}
The effects of pressure on a quantum spin liquid are investigated in an organic Mott insulator $\kappa$-(ET)$_2$Ag$_2$(CN)$_3$ with a spin-1/2 triangular lattice. The application of negative chemical pressure to $\kappa$-(ET)$_2$Cu$_2$(CN)$_3$, which is a well-known sister Mott insulator, allows for extensive tuning of antiferromagnetic exchange coupling, with $J/k_{\rm B} = 175 - 310$ K, under hydrostatic pressure. Based on $^{13}$C nuclear magnetic resonance measurements under pressure, we uncover universal scaling in the static and dynamic spin susceptibilities down to low temperatures $\sim 0.1k_{\rm B}T/J$. The persistent fluctuations and residual specific heat coefficient are consistent with the presence of gapless low-lying excitations. Our results thus demonstrate fundamental finite-temperature properties of quantum spin liquid in a wide parameter range. 
\end{abstract}



\maketitle

Quantum spin liquids featuring collectively entangled spin-singlet pairs possess emergent low-energy excitations or quasiparticles involving fractionalized fermionic spinons and topological vortices termed as visons \cite{Anderson, Balents, Alicea, Lee, Wen}. Identification of such excitations has been an experimental challenge for decades. In contrast to classical liquids (e.g., water) and Fermi liquids (e.g., metal and $^3$He), which are well characterized in terms of density or interaction strength, systematic studies on quantum spin liquids have not been carried out, despite the abundance of candidate materials reported so far, including triangular lattice organics \cite{Shimizu1, Pratt, Itou, Isono} and kagome lattice cuprates \cite{Mendels, Helton}. 

For a triangular lattice, a spin liquid state appears in a Mott insulator with moderate electron correlation $U/t$ or ring exchange coupling, where $U$ denotes the on-site Coulomb repulsion and $t$ denotes the transfer integral. Indeed, long-range magnetic ordering is absent in a molecular Mott insulator $\kappa$-(ET)$_2$Cu$_2$(CN)$_3$ [ET denotes bis(ethylenedithio)tetrathiafulvalene] with a triangular lattice of molecular dimers (a ratio of transfer integrals between dimers, $t^\prime/t \sim 1$, and $U/t \sim 8$) \cite{Shimizu1, Pratt}, located close to a superconducting phase across the insulator-to-metal (Mott) transition under hydrostatic pressure \cite{Kurosaki}. The phase diagram is distinct from those of less-frustrated antiferromagnets in which conventional N${\rm \acute{e}}$el order phases reside near the superconducting phases \cite{Uemura}. The pairing mechanism of the superconducting phase appearing from the spin liquid may involve emergent features such as triplet \cite{Lee, Galitski} and excitonic pairing \cite{Qi2}. Low-lying excitations of the spin liquid state are investigated by heat capacity \cite{YamashitaS} and optical conductivity \cite{Dressel, Kezsmarki}, supporting a gapless spinon picture \cite{Motrunich, Lee, Lee3, Galitski, Zou}, while magnetic resonances \cite{Pratt, Shimizu2} are compatible with a spin liquid with a small/nodal gap \cite{Alicea, Qi, Mishmash}. The excitations are also discussed in terms of residual internal charge/lattice degrees of freedom \cite{Naka, Watanabe} based on the ultrasound \cite{Poirier}, thermal expansion \cite{Manna}, and dielectric measurements \cite{Jawad, Itoh, Dressel2}.  

As the system becomes more insulating, it is known that a Heisenberg model with nearest neighbor interactions gives a spiral long-range order \cite{Huse, Bernu}, requiring a quantum phase transition from the spin liquid at a larger $U/t$ (= $9-13$) in the Hubbard model \cite{Mizusaki, Yoshioka, Sahebsara}. Such a strong Mott insulator has not been developed in the series of organic compounds $\kappa$-(ET)$_2$X (X: anions). The realization of such a Mott insulator can facilitate systematic studies of the quantum spin liquid through continuous tuning of electron correlation and geometrical frustration. 

In this Letter, for pushing the spin liquid phase into a larger $U/t$ region, we report a newly synthesized compound $\kappa$-(ET)$_2$Ag$_2$(CN)$_3$ with an expanded triangular lattice ($t^\prime /t = 0.97$, $U/t = 9.2$). We have succeeded in applying a negative chemical pressure to develop the strong Mott insulator. It allows an extensive variation of $U/t$ or exchange coupling $J$ with hydrostatic pressure, maintaining the system as an insulator without long-range magnetic order. Constructing the pressure-temperature ($P$-$T$) diagram based on resistivity and nuclear magnetic resonance (NMR) measurements, we investigate the local spin susceptibility as functions of temperature, pressure, and magnetic field. Together with specific heat results, we discuss low-lying excitations that persist toward low temperatures and the Mott boundary. 

Single crystals of $\kappa$-(ET)$_2$Ag$_2$(CN)$_3$ were grown by electro-oxidation of ET with KAg(CN)$_2$ and 18-crown-6 in 1,1,2-trichloroethane/ethanol \cite{Hiramatsu}. Single-crystal X-ray diffraction measurements show the crystal structure [Fig. 1(a), Supplemental Material Fig. S1] of monoclinic space group $P2_1/c$ with lattice constants $a$ = 15.055(1), $b$ = 8.7030(7), $c$ = 13.412(1)$\AA$, $\beta$ = 91.307(1)$^\circ$, and $V$ = 1756.8(2)$\AA^3$ at 300 K \cite{SP}. The interlayer resistivity and $^{13}$C NMR measurements were carried out on a single crystal at ambient and hydrostatic pressures. A polycrystalline sample was used for the magnetization, specific heat, and $^{1,2}$H NMR measurements. For $^{13}$C NMR, $^{13}$C isotope ($>$ 94 atom $\%$) was selectively enriched on central double-bonded carbon sites of ET \cite{Larsen}. The $^{13}$C NMR spectra and the nuclear spin-lattice relaxation rate $T_1^{-1}$ were obtained using Fourier transformed spin-echo signals in a magnetic field parallel to the conducting layers (the $c$ axis). Superconducting transition was observed using an NMR coil at zero field but was suppressed under a magnetic field of 8.5 T, which exceeds the critical field due to misalignment ($< 2^\circ$) from the vortex lock-in direction. 

	\begin{figure}
	\includegraphics[scale=0.65]{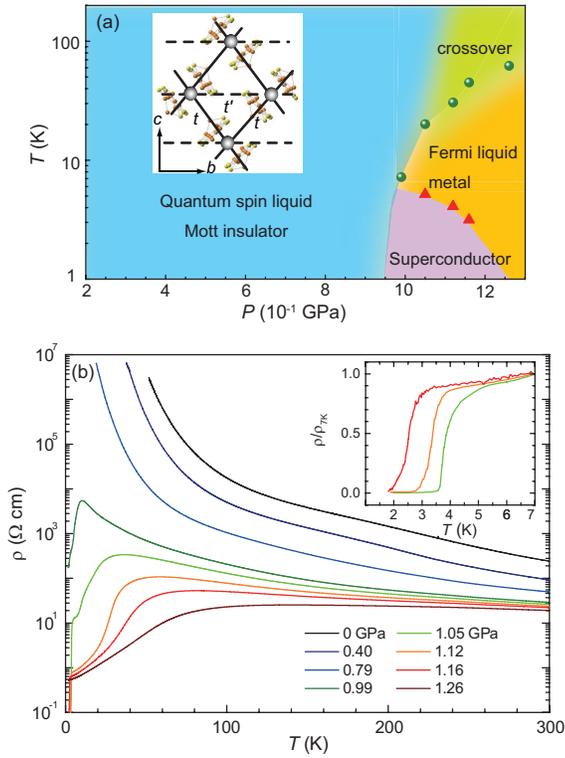}
	\caption{\label{Fig1} 
(a) Pressure-temperature phase diagram of $\kappa$-(ET)$_2$Ag$_2$(CN)$_3$. The circle and triangle symbols denote Mott and superconducting transition temperatures, respectively, determined by (b) resistivity $\rho$ measurements for various pressures. Inset: (a) triangular lattice of ET dimers, (b) low-temperature resistivity normalized at 7 K for 1.05-1.16 GPa. 
	}
	\end{figure}
First, we construct the $P$-$T$ phase diagram of $\kappa$-(ET)$_2$Ag$_2$(CN)$_3$ from the temperature dependence of resistivity $\rho$ under hydrostatic pressures (Fig. 1). At ambient pressure, $\rho$ exhibits as an insulator with a charge gap $\Delta_c/k_{\rm B} = 1.2 \times 10^3$ K ($k_{\rm B}$: the Boltzmann constant) at 300 K (Fig. S2) \cite{SP}. $\Delta_c$ decreases with increasing pressure, and Mott transition occurs at $T_{\rm IM}$ = $20$ K (defined as the resistivity inflection point, $d \rho/d T$ maximum) for $P_c = $ 1.05 GPa. $T_{\rm IM}$ increases with increasing pressure, showing a positive slope of the phase boundary, $d T_{\rm IM}/d P > 0$. Superconducting transition is observed at the onset temperature $T_{\rm SC}$ = 5.2 K ($P = P_c$) and is suppressed at a rate of $dT_{\rm SC}/dP = -1.8 {\rm K}/10^{-1}$ GPa. The phase diagram is similar to that of $\kappa$-(ET)$_2$Cu$_2$(CN)$_3$ \cite{Kurosaki} except for a pressure offset $P_0$ = $0.6$ GPa, which quantifies the negative chemical pressure by the Ag substitution. 

	\begin{figure}
	\includegraphics[scale=0.67]{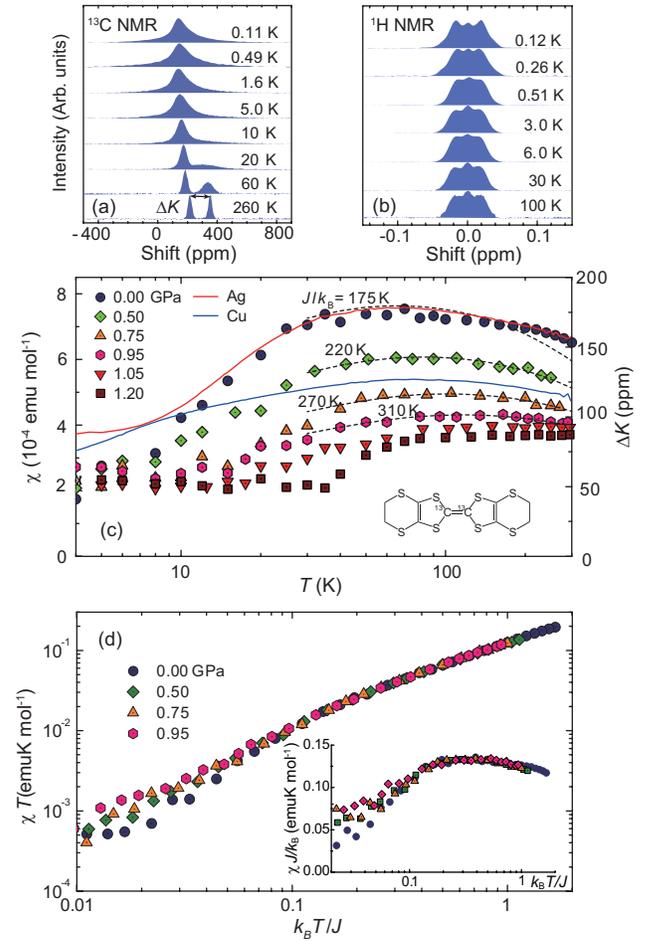}
	\caption{\label{Fig2} 
(a) $^{13}$C (8.5 T) and (b) $^1$H (2.0 T) NMR spectra at ambient pressure, where the horizontal axis is defined as $(\omega - \omega_0)/\omega_0$ ($\omega_0$: the resonance frequency of tetramethylsilane). (c) Spin susceptibility $\chi$ obtained from the magnetization (red solid curve) at ambient pressure and from the relative $^{13}$C Knight shift $\Delta K$ (solid symbols) for various pressures. Dotted curves are fitted to a triangular-lattice Heisenberg model $\mathcal{H} = \sum_{i,j} J {\bf S}_i \cdot {\bf S}_j$. The blue solid curve is $\chi$ of the Cu salt \cite{Shimizu1}. Inset: $^{13}$C enriched ET molecule. (d) $\chi T$ versus $k_{\rm B}T/J$ plots for $P < P_c$. Inset: $\chi J/k_{\rm B}$ versus $k_{\rm B}T/J$ plots. 
	}
	\end{figure}	
The spin susceptibility $\chi$ was obtained from magnetization and $^{13}$C NMR [Fig. 2(a)] measurements. As shown in Fig. 2(c), a weak temperature dependence of $\chi$ with a broad maximum would be characteristic of frustrated antiferromagnets in two dimensions. $\chi$ is fitted to a triangular-lattice Heisenberg model \cite{Elstner} with the exchange coupling $J/k_{\rm B}$ = 175 K at ambient pressure. A slight difference from the model may arise from the thermal lattice contraction. $\chi$ decreases steeply below 30 K, reflecting short-range spin correlations. Nevertheless, $\chi$ remains finite down to low temperatures ($> 1.9$ K). Similar behavior is observed in the local spin susceptibility obtained from the $^{13}$C Knight shift without including a Curie impurity contribution [Fig. 2(c)]. 

The $^1$H NMR technique is utilized to probe spontaneous local moments \cite{Miyagawa}. As seen from Fig. 2(b), the shape of the spectrum is governed by nuclear dipole coupling down to 0.12 K, whereas a slight change below 0.26 K is due to an extrinsic origin (Fig. S4). The $T$-invariant linewidth rules out a magnetic moment larger than $0.01\mu_{\rm B}$ ($\mu_{\rm B}$: the Bohr magneton) at 2.0 T. Moreover, we confirmed the absence of structural distortions by x-ray diffraction measurements above 8 K \cite{Hiramatsu}. These results strongly suggest that a quantum disordered or liquid state persists for $T < 0.001J/k_{\rm B}$. Instead of spontaneous moments, the $^{13}$C NMR spectra at 8.5 T show a site and field-dependent broadening below 30 K [Figs. 2(a), S5, S6], which are attributable to field-induced staggered moment with the amplitude less than 0.02$\mu_{\rm B}$ \cite{Shimizu2, SP}.

The susceptibility $\chi$ under pressure is obtained from the $^{13}$C Knight shift measurements by using a hyperfine coupling constant $0.13$ T/$\mu_{\rm B}$ at ambient pressure. As shown in Fig. 2(c), $\chi$ decreases with increasing $P$. The temperature dependences conform well with the triangular-lattice Heisenberg model \cite{Elstner} with $J/k_{\rm B}$ = 220, 270, and 310 K, for $P$ = 0.5, 0.75, and 0.95 GPa, respectively. Note that $\chi$ of $\kappa$-(ET)$_2$Cu$_2$(CN)$_3$ ($J/k_{\rm B}$ = 250 K) \cite{Shimizu1} is located in between those of 0.50 and 0.75 GPa [blue solid curve in Fig. 2(c)], in agreement with the effective chemical pressure (0.6 GPa) evaluated from the critical pressure of the Mott transition. 

In the Heisenberg model, thermodynamic properties are governed by $J$ in paramagnetic states. $\chi$ would be expressed as a function of $x = k_{\rm B}T/J$. As shown in the $\chi T$ versus $k_{\rm B}T/J$ plot [Fig. 2(d)], we found an excellent scaling property down to low temperatures, $k_{\rm B}T/J < 0.1$, whereas a high-$T$ series expansion is valid only for $k_{\rm B}T/J > 0.2$ \cite{Elstner}. Thus, $\chi$ is expressed as $T^{-1}f(x)$, where $f(x)$ denotes a scaling function of $x$. The result justifies the fitting with the triangular-lattice Heisenberg model up to the Mott boundary. If $\chi$ at low temperatures ($T \ll J/k_{\rm B}$) is dominated by fermionic spinons, $\chi$ can be proportional to spinon density of states $n_{\rm s}$ and inversely scales with the spinon hopping amplitude governed by $J$ ($n_{\rm s} \propto  J^{-1}$) in a weakly correlated model \cite{Motrunich}. In fact, $\chi$ is approximately proportional to $J^{-1}$ [inset of Fig. 2(d)] for $k_{\rm B}T/J > 0.1$ and approaches a pressure-insensitive value ($2-3 \times 10^{-4}$ emu mol$^{-1}$) comparable to the Pauli paramagnetic susceptibility in the metallic state above $P_c$. 

	\begin{figure}
	\includegraphics[scale=0.73]{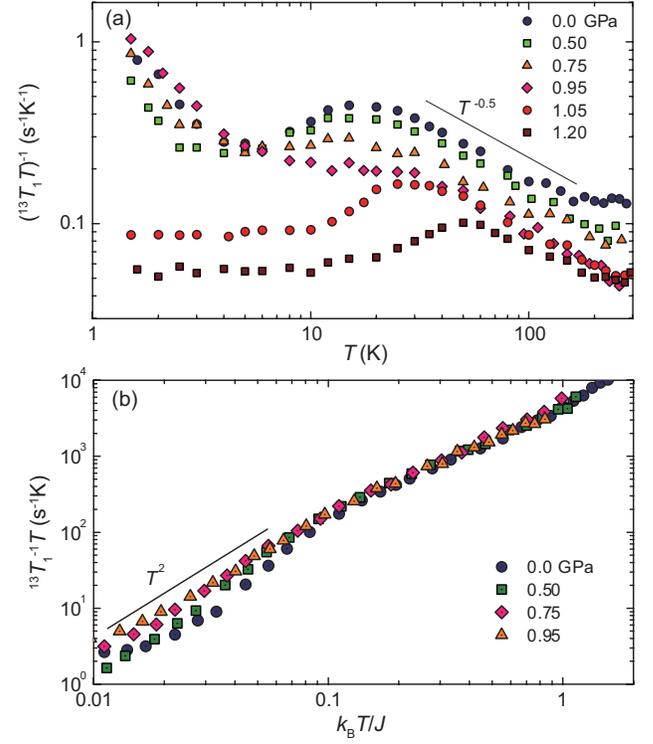}
	\caption{\label{Fig3} 
	(a) Temperature dependence of nuclear spin-lattice relaxation rate divided by temperature, $(^{13}T_1T)^{-1}$, and (b) $^{13}T_1^{-1}T$ versus $k_{\rm B}T/J$ plots for $P < P_c$ in $\kappa$-(ET)$_2$Ag$_2$(CN)$_3$. 
	}
	\end{figure}
Low-lying spin excitations can be more sensitively probed via the nuclear spin-lattice relaxation rate $T_1^{-1}$. In paramagnetic systems, $T_1^{-1}$ is given by \cite{Moriya} 
	\begin{eqnarray}
	T_1^{-1} = \frac{2\gamma_{\rm n}^2 k_{\rm B}T}{\gamma_{\rm e}^2\hbar}\sum_{\bf q} A_{\bf q}A_{-\bf q}\frac{\chi^{\prime \prime}({\bf q}, \omega_{\rm n})}{\omega_{\rm n} } 
	\end{eqnarray} 
with the nuclear (electron) gyromagnetic ratio $\gamma_{\rm n} (\gamma_{\rm e})$, reduced Planck's constant $\hbar$, hyperfine form factors $A_{\bf q}A_{\bf -q}$, and the imaginary part of dynamical spin susceptibility $\chi^{\prime \prime}({\bf q}, \omega_{\rm n})$ at wave vectors {\bf q} and NMR frequency $\omega_{\rm n}$. As shown in Fig. 3(a), $(^{13}T_1T)^{-1}$ increases upon cooling ($\sim T^{-0.5}$) owing to the weak evolution of the antiferromagnetic correlation length $\xi$, which can be characteristic of triangular-lattice quantum antiferromagnets \cite{Elstner}. Based on the dynamical scaling argument, the relation, $(T_1T)^{-1} \propto \xi^{2+z - \eta-d} \simeq T^{-0.5}$, gives $z-\eta \approx  0.5$ in two dimensions ($d=2$), where $z$ and $\eta$ are the dynamical and critical exponents, respectively. This is in contrast to the quantum critical behavior ($z-\eta =1$) in less frustrated antiferromagnets \cite{Chubukov}. $(^{13}T_1T)^{-1}$ levels off below 30 K and tends to grow below $5-6$ K, as discussed below. 

With increasing pressure $(^{13}T_1T)^{-1}$ is suppressed and exhibits a sudden decrease at $T_{\rm IM}$ ($T_{\rm IM} = 20$ K at 1.05 GPa and 50 K at 1.20 GPa), consistent with the resistivity result. It signifies the reduction of $\chi^{\prime \prime}({\bf q}, \omega_{\rm n})$ over the ${\bf q}$ space with increasing $P$ or reducing $U/t$. Below $T_{\rm IM}$, the system enters into a Fermi liquid state with the Korringa's law, $(^{13}T_1T)^{-1}$ = constant. $T_1^{-1}$ is expected to scale to $T^\eta \Phi (x)$ with a scaling function $\Phi (x)$ in the quantum critical or disordered regime \cite{Chubukov}. Indeed, we found the scaling property for $\eta \sim 1$ as a function of $x = k_BT/J (<1)$ in Fig. 3(b). Below $0.1k_{\rm B}T/J$, $^{13}T_1^{-1}T$ approaches $\sim T^2$ behavior, corresponding to the Korringa's law in the spin liquid with $z - \eta = 0$ ($z \sim 1$). The result supports the predominant gapless fermionic excitations down to $k_{\rm B}T/J < 0.1$. A deviation of the scaling below $0.1k_{\rm B}T/J$ suggests growing contributions of the gapped excitations such as preformed spinon pairing \cite{Nave} and the fractionalized vison excitations with $\eta = 1.37$ in a critical regime \cite{Qi}. The reduction of $^{13}T_1^{-1}$ at lower pressures seems incompatible with the Hubbard model predicting a quantum phase transition into a spiral phase \cite{Mizusaki, Yoshioka, Sahebsara}. The absence of the critical behavior is justified, if the transition into the spiral phase is first order \cite{Yoshioka}. 

	\begin{figure}
	\includegraphics[scale=0.62]{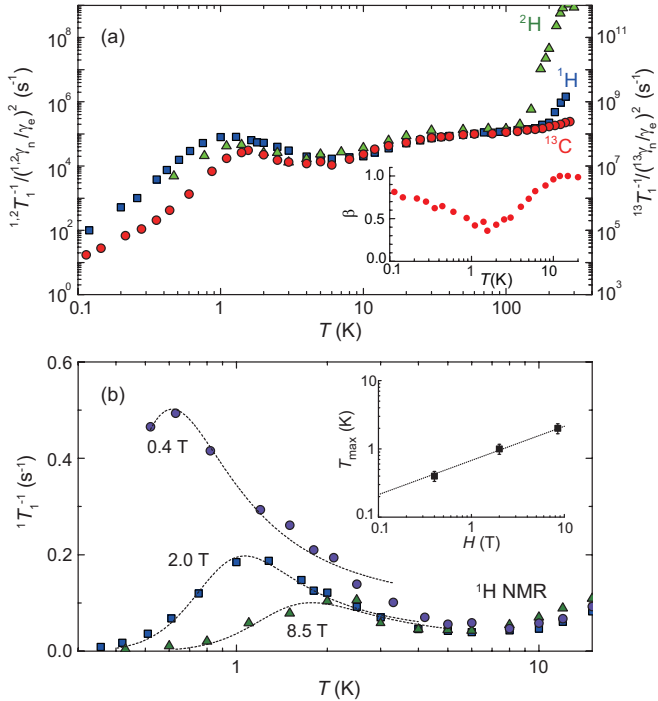}
	\caption{\label{Fig4} 
(a) Nuclear spin-lattice relaxation rates $^iT_1^{-1}$ normalized with respect to square of gyromagnetic ratios, $^i\gamma_{\rm n}^2/\gamma_e^2$ ($i = 1$, $2$, and $13$), for 2.0 T ($^1$H) and 8.5 T ($^2$H and $^{13}$C). Inset: stretched exponent $\beta$ of the $^{13}$C magnetization recovery. (b) $^1T_1^{-1}$ measured at several magnetic fields. Dotted curves are fitted to Lorentzian-type fluctuations. Inset: field dependence of $T_{\rm max}$, $^1T_1^{-1}$ peak temperature. 
	}
	\end{figure}
It would be crucial to specify the origin of anomalous low-temperature fluctuations in the insulating phase, which may be related to fractionalization of spin liquid quasiparticles and to superconducting pairing under pressure. We compare $^iT_1^{-1}$ normalized with respect to $(^i\gamma_{\rm n}/\gamma_{\rm e})^2$ ($i$ = $1, 2$, and $13$) for $^1$H, $^2$H and $^{13}$C sites with different hyperfine and quadrupole interactions at ambient pressure [Fig. 4(a)]. As observed above 200 K, $^2T_1^{-1}$ more sensitively detects thermal molecular motions than $^1T_1^{-1}$. The linear scaling among $^iT_1^{-1}$ ($i$ = $1, 2$, and $13$) below 200 K confirms predominant spin fluctuations following Eq. (1). Below 6 K, $^iT_1^{-1}$ again weakly depends on the nuclear sites with a broad maximum around $T_{\rm max} = 1-2$ K and then follows $T^2$ dependence toward $T = 0$. Here $^{13}T_1^{-1}$ includes several components, as manifested in a decrease of the exponent $\beta$ for a stretched exponential fit to the nuclear magnetization recovery $M(t) = A$exp[$-(t/^{13}T_1)^\beta$] at low temperatures [inset of Fig. 4(a), Fig. S7], whereas $^1T_1^{-1}$ and $^2T_1^{-1}$ consist of a single component. The difference indicates the existence of microscopic heterogeneity over several unit cells, which is averaged for the $^1$H sites via nuclear spin-spin relaxation much faster than the $T_1$ process. 

As shown in Fig. 4(b), $^1T_1^{-1}$ exhibits remarkable field dependence in the field range, 0.4-8.5 T. The temperature dependence conforms to Lorentzian type fluctuations with a characteristic correlation time $\tau_c$ \cite{BPP, SP} [dotted curves in Fig. 4(b), Supplemental Material]. $T_{\rm max}$ decreases with $\sim H^{0.5}$, pointing to extremely slow fluctuations persistent in the zero field limit. One can eliminate the possibility of a spin glass state because the NMR spectrum shows no change across $T_{\rm max}$. The field sensitivity is reminiscent of the critical behavior observed in $\mu$SR measurements in $\kappa$-(ET)$_2$Cu$_2$(CN)$_3$ \cite{Pratt}. 

	\begin{figure}
	\includegraphics[scale=0.6]{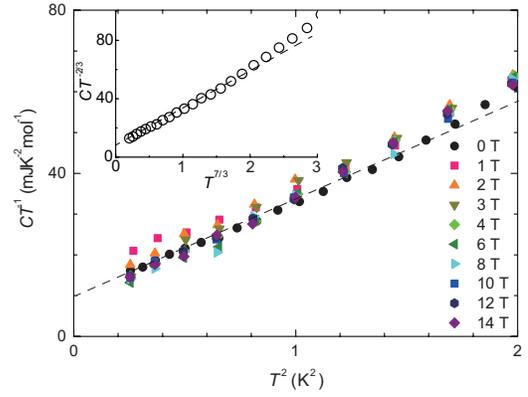}
	\caption{\label{Fig5} 
Squared temperature dependence of specific heat divided by $T$ for magnetic field in the range 0-14 T in $\kappa$-(ET)$_2$Ag$_2$(CN)$_3$. Inset shows $CT^{-2/3}$ versus $T^{7/3}$ plot at $ H = 0$ T. 
	}
	\end{figure}

To examine a possible field-induced transition, the field dependence of specific heat $C$ was measured at low temperatures, as shown in Fig. 5. We find no appreciable field dependence in $CT^{-1}$ plotted as a function of $T^2$. The result is incompatible with the field-induced magnetic ordering often encountered in spin-dimer systems \cite{Brown}. The observed field-dependence in $^1T_1^{-1}$ thus should be of purely dynamical origin. This type of slow dynamics may be caused by valence bond fluctuations coupled to phonons, and thus be suppressed when the field-induced staggered moments grow at higher fields. It is noted that the $CT^{-1}$ intercept known as the $\gamma$ term remains finite ($\gamma = 10$ mJK$^{-2}$ mol$^{-1}$ at zero field). It agrees with the low-lying property of fermionic spinons ($C \sim T^n$, $n = 2/3 - 1$) \cite{Wang}, as also shown in the inset of Fig. 5. The obtained $\gamma$ is slightly lesser than that of the Cu compound ($\gamma = 12$ mJK$^{-2}$ mol$^{-1}$) \cite{YamashitaS}. It suggests that the spinon effective mass is nearly independent of $J$, consistent with the pressure-insensitive $\chi$ and $T_1^{-1}$ at low temperatures. 

The present compound has several prominent features for studying quantum spin liquid systematically. (1) As shown above, the application of negative pressure to the sister compound $\kappa$-(ET)$_2$Cu$_2$(CN)$_3$ enables extensive tuning of exchange coupling and electron correlations, while maintaining the triangular lattice anisotropy $t^\prime/t$ close to unity. It is also accessible to the superconducting phase across the Mott transition. (2) The Ag substitution eliminates a sample dependence problem encountered in the Cu compound having paramagnetic Cu$^{2+}$ impurities. (3) The distance between ET and anion is shorter for the Ag salt than the Cu salt. This can affect the ethylene group conformation through the hydrogen bond interactions with anion layers. Unlike the other $\kappa$-(ET)$_2$X compounds, the ordering of ethylene groups from 300 K results in weaker inhomogeneous local fields as observed in the $^{13}$C NMR spectrum ($<0.02\mu_{\rm B}$ at 8.5 T). Instead, the Dzyaloshinskii-Moriya interaction is relevant to the field-induced moments at low temperatures \cite{DM}, since the C-N bond disorder in the anion layer locally breaks the inversion. 

In conclusion, we have demonstrated that a newly synthesized organic Mott insulator $\kappa$-(ET)$_2$Ag$_2$(CN)$_3$ with a triangular lattice possesses extended quantum spin liquid phase close to the superconducting phase. The application of negative chemical pressure in well studied $\kappa$-(ET)$_2$Cu$_2$(CN)$_3$ pushes the system away from the Mott transition, which allows the systematic investigation of low-lying excitations as functions of pressure. We have uncovered the universal scaling in the static and dynamic susceptibilities, and the presence of fractional excitations. The result highlights the fundamental thermodynamic properties of quantum spin liquid on the triangular lattice Mott insulator. 

We thank M. Tsuchiizu, C. Hotta, A. Kobayashi, M. Imada, H. Kishida, and H. Ito for fruitful discussions. This work was supported by JSPS KAKENHI Grant Numbers JP23225005 and JP26288035, and the Visiting Researcher's Program of the Institute for Solid State Physics, University of Tokyo. 


\end{document}